\documentclass[fleqn,twoside]{article}
\usepackage{espcrc2}
\usepackage{graphicx}
\usepackage[figuresright]{rotating}
\newcommand{\BR}{{\cal B}_0}

\newcommand{\psip}{\psi(2S)}

\newcommand{\EE}{e^+e^-}
\newcommand{\MM}{\mu^+\mu^-}

\newcommand{\PP}{\pi^+\pi^-}

\newcommand{\OP}{\omega\pi^0}

\newcommand{\beq}{\begin{equation}}
\newcommand{\eeq}{\end{equation}}
\newcommand{\beqn}{\begin{eqnarray}}
\newcommand{\eeqn}{\end{eqnarray}}
\newcommand{\beqns}{\begin{eqnarray*}}
\newcommand{\eeqns}{\end{eqnarray*}}
\newcommand{\bfg}{\begin{figure}}
\newcommand{\efg}{\end{figure}}
\newcommand{\bitm}{\begin{itemize}}
\newcommand{\eitm}{\end{itemize}}
\newcommand{\bnum}{\begin{enumerate}}
\newcommand{\enum}{\end{enumerate}}
\newcommand{\btbl}{\begin{table}}
\newcommand{\etbl}{\end{table}}
\newcommand{\btbu}{\begin{tabular}}
\newcommand{\etbu}{\end{tabular}}

\title{The form factors of $\omega\pi^0$ and $\pi^+\pi^-$ 
at $\psi(2S)$} 
\author{P.~Wang \address[IHEP]{Institute of High Energy Physics,
P.O.Box 918, Beijing 100039, China}
\thanks{Supported by National Natural Science Foundation of China (19991483)
and 100 Talents Program of CAS (U-25)},
X.~H.~Mo\addressmark[IHEP]$^,$\address[CCAST]{China Center of Advanced Science and
Technology, Beijing 100080, China},
C.~Z.~Yuan \addressmark[IHEP] }

\date{}
\begin{document}

\begin{abstract}
The measurements of $\psi(2S)\rightarrow\OP$ and
$\psi(2S)\rightarrow \PP$ in $\EE$ experiments are
examined. It is found that the non-resonance virtual
photon annihilation gives large contributions to the
observed cross sections of these two processes.
By including this contribution, the form factors and
branching fractions of these two decay modes are revised.
\vspace{1pc}
\end{abstract}
\maketitle

\section{Introduction}
Since its discovery, a large amount of $\psi(2S)$ data has 
been collected. 
The latest comes from BES~\cite{bes}. This has led to detailed
analysis of the interference pattern between the strong and 
the electromagnetic interactions in $\psip$ decays~\cite{suzuki}. 
In such analysis, the electromagnetic decay modes such as 
$\OP$ and $\PP$ are of particular importance~\cite{suzuki,chernyak}.   

\begin{figure}
\begin{minipage}{8cm}
\includegraphics[width=3.25cm,height=2.5cm]{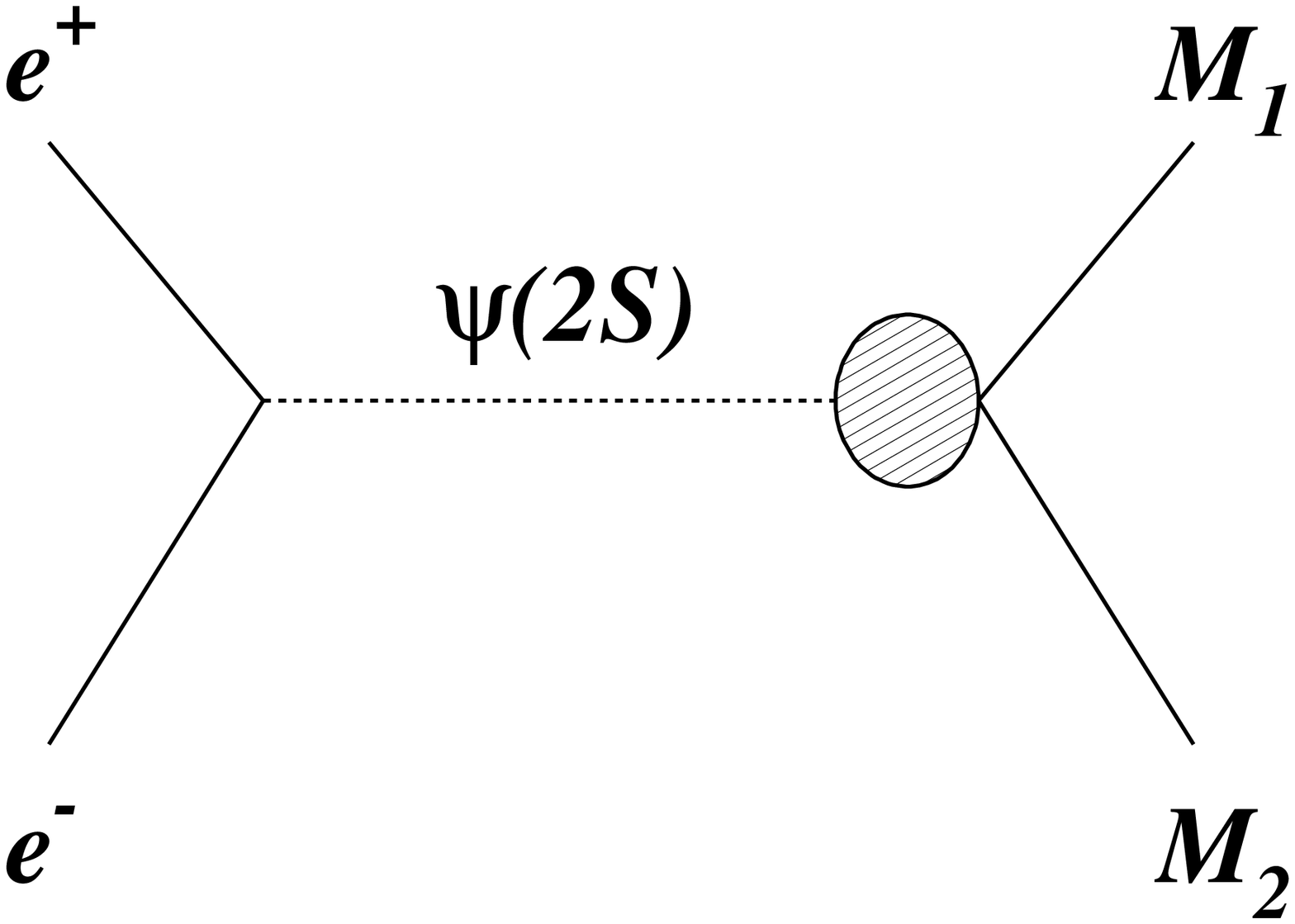}
\hskip 0.5cm
\includegraphics[width=3.25cm,height=2.5cm]{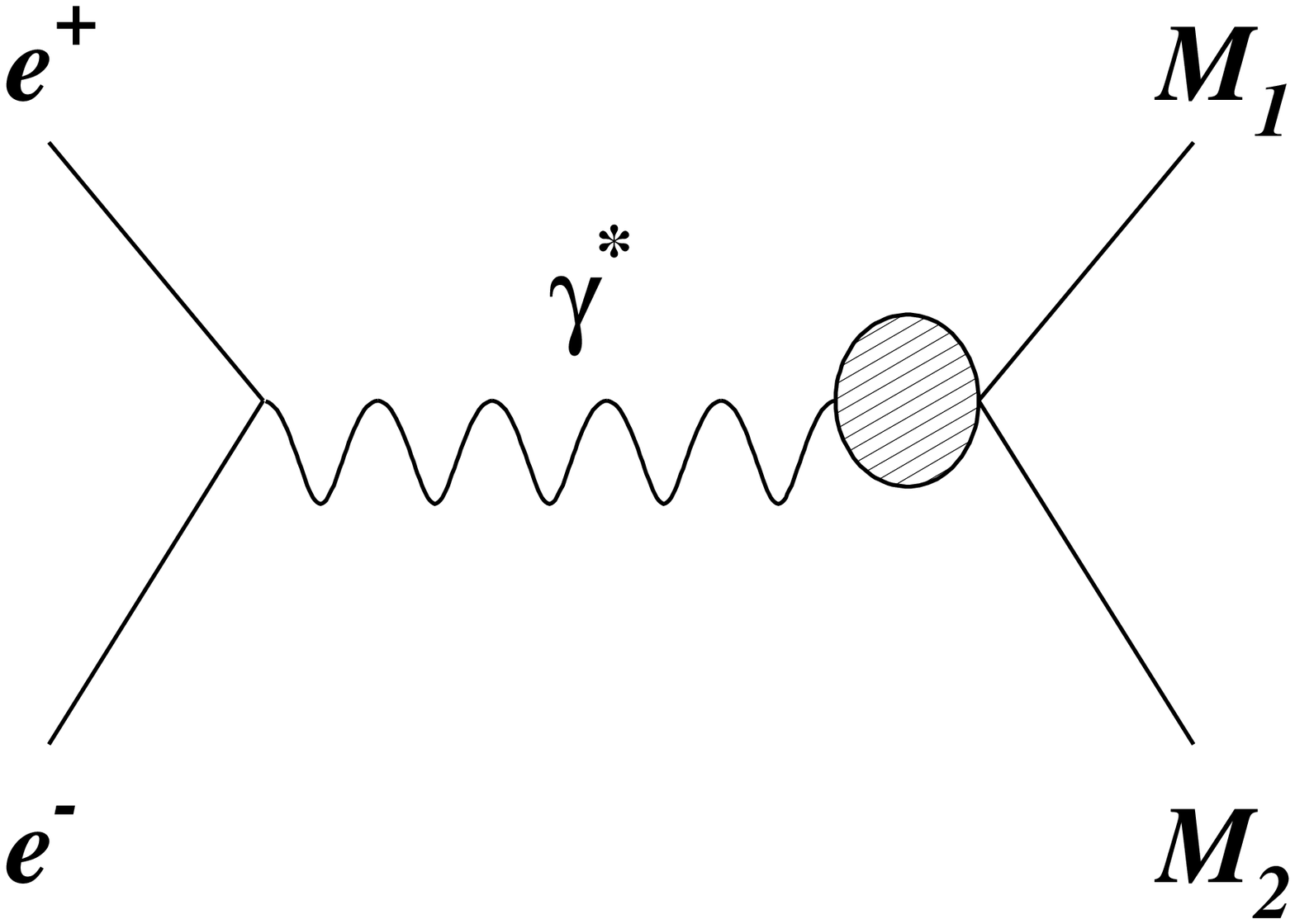}
\end{minipage}
\caption{\label{feymnfig} Feynman diagrams for
$\EE\rightarrow M_1 + M_2$ near $\psip$.}
\end{figure}

Up to now, the most precise measurements of the $\psi(2S)$ decays  
are by $e^+e^-$ colliding experiments, where the production of 
$\psi(2S)$ is accompanied by 
$$
e^+e^- \rightarrow \gamma^* \rightarrow hadrons ,
$$
in which $e^+e^-$ pair annihilates into a virtual photon without 
going through the intermediate resonance state. 
So the experimentally measured $\psip \rightarrow \OP$
and $\psip \rightarrow \pi^+\pi^-$ processes are parallel to 
$\psip \rightarrow \mu^+\mu^-$ in the way that there are
two Feynman diagrams: one is the $\psi(2S)$ and the other
is the one-photon annihilation, as shown in Fig.~\ref{feymnfig}.
There are three terms in the cross
section: $\psi(2S)$ resonance, one-photon annihilation
and their interference. Unlike $\mu^+\mu^-$ pair which couples 
to virtual photon by QED fine structure constant, the couplings 
of $\OP$ and $\PP$ to virtual photon are by energy-dependent 
form factors which are to be determined by experiments.
By scanning around the $\psip$ peak, $\Gamma_{ee}$ and $\Gamma_{\mu\mu}$
are determined from fitting the resonance shape with the theoretical 
curve, which includes the one-photon annihilation term
and interference~\cite{scan}. 
On the contrary, $\Gamma_{\OP}$ and $\Gamma_{\PP}$, 
due to their small branching fractions,  
are acquired from data collected on top of 
the resonance. The contributions from the one-photon annihilation 
and the interference are not subtracted.

In this work, the measurements of the $\OP$
and $\PP$ form factors at $\psip$ resonance by $\EE$ 
colliding experiments are examined. First, not only the $\psip$ resonance,
but also the one-photon annihilation propagator are included in 
the experimentally observed cross section, which takes into account
the initial state radiation and the finite energy 
resolution of the $\EE$ colliders.
Next it is demonstrated that for current measurements,
the one-photon annihilation contributes a large share
to the observed cross sections of these two processes, 
and this leads to significant revision of the form 
factors of $\OP$ and $\PP$ and their branching fractions
in $\psip$ decays.
Finally the dependence of the measurement on the experimental 
condition is discussed in detail.
 
\section{The experimentally observed cross section} 
The resonance part of the cross sections for 
$\EE\rightarrow\psip\rightarrow\OP$ and 
$\EE\rightarrow\psip\rightarrow\pi^+\pi^-$,
in the Born order, are expressed by the Breit-Wigner formula
$$
\sigma_{Born}(s)\;=\frac{12\pi \Gamma_{ee} \Gamma_{f}}{(s-M^{2})^{2}
+\Gamma^{2}_t M^{2}}. 
$$
Here $\sqrt{s}$ is the center of mass energy; 
$M$ and $\Gamma_t$ are the mass and the total width of $\psip$;
$\Gamma_{ee}$ is the partial width to $\EE$, and 
$\Gamma_{f}$ ($f=\OP$, $\PP$) is the partial width to the final state $f$, 
which are related to $\Gamma_{ee}$ and the corresponding form factors :
\beq
\Gamma_{\OP} =  \frac{\Gamma_{ee}
q^3_{\OP}}{m_{\psip}} |{\cal F}_{\OP}(m^2_{\psip})|^2,
\label{FOP}
\eeq
and
\beq
\Gamma_{\PP} = 2\Gamma_{ee}\left(\frac{q_{\PP}}{m_{\psip}}\right)^3 
|{\cal F}_{\PP}(m^2_{\psip})|^2.
\label{FPP}
\eeq
Here $q_{\OP}$ is the momentum of either $\omega$ or $\pi^0$
in the $\OP$ decay, $q_{\PP}$ is the momentum of $\pi$ in the 
$\PP$ decay.  ${\cal F}_{\OP}(s)$ and ${\cal F}_{\PP}(s)$ 
are the form factors of $\OP$ and $\PP$, respectively. 

For the experimentally observed cross sections of $\EE\rightarrow\OP$ 
and $\EE\rightarrow\PP$ at $\psip$ peak, the direct one-photon 
annihilation term and an interference term must be added:
\beqn
\lefteqn{\sigma_{Born}(s)= } \label{Born} \\
& & \frac{4\pi\alpha^2}{s^{3/2}}[1+2\Re B(s)+|B(s)|^2]
|{\cal F}_f(s)|^2{\cal P}_f(s), \nonumber
\eeqn
with
\beq
B(s)=\frac{3\sqrt{s}\Gamma_{ee}/\alpha}{s-M^2+iM\Gamma_t},
\label{bexpr}
\eeq
where $\alpha$ is the QED fine structure constant, and 
$$
{\cal P}_{\OP}(s) = \frac{1}{3}q^3_{\OP},
$$
$$
{\cal P}_{\PP}(s) = \frac{2}{3s}q^3_{\PP}.
$$
More generally, there could be a 
phase between the one-photon and $\psip$ propagators,
then instead of Eq.~(\ref{bexpr}), one has
\beq
B(s)=\frac{3\sqrt{s}\Gamma_{ee}/\alpha}{s-M^2+iM\Gamma_t}e^{i\phi},
\label{phi}
\eeq
where $\phi$ is the phase between the two propagators. In the 
following analysis, both Eq.~(\ref{bexpr}) and (\ref{phi}) are
considered.

In $\EE$ collision, the Born order cross section is
modified by the initial state radiation in the way~\cite{rad.1} 
\begin{equation}
\sigma_{r.c.} (s)=\int \limits_{0}^{x_m} dx 
F(x,s) \frac{\sigma_{Born}(s(1-x))}{|1-\Pi (s(1-x))|^2},
\label{RAD}
\end{equation}    
where $x_m=1-s'/s$.
$F(x,s)$ has been calculated to an accuracy of 
0.1\%~\cite{rad.1,rad.2,rad.3}
and $\Pi(s)$ is the vacuum polarization factor.
In the upper limit of the integration, $\sqrt{s'}$ 
is the experimentally required 
minimum invariant mass of the final particles.
In this work, $x_m=0.2$ is used which corresponds to
invariant mass cut of 3.3~GeV.

By convention, $\Gamma_{ee}$ has the QED vacuum polarization in 
its definition~\cite{Tsai,Luth}. Here it is natural to 
extend this convention to the partial widths of other pure 
electromagnetic decays. 
By using Eq.~(\ref{FOP}) and (\ref{FPP}) to relate
$\Gamma_{\OP}$ and $\Gamma_{\PP}$ with $\Gamma_{ee}$, 
these partial widths are experimentally measured ones
with vacuum polarization implicitly included in their definitions.

The $\EE$ colliders have finite energy resolution 
which is much wider than the intrinsic width of $\psip$. 
Such energy resolution is usually a Gaussian distribution:
$$
G(W,W^{\prime})=\frac{1}{\sqrt{2 \pi} \Delta}
             e^{ -\frac{(W-W^{\prime})^2}{2 {\Delta}^2} },
$$
where $W=\sqrt{s}$ and $\Delta$, a function of the energy, 
is the standard deviation of the Gaussian distribution. 
The experimentally measured cross section is the
radiative corrected cross section folded with the energy 
resolution function
\begin{equation}
\sigma_{exp} (W)=\int \limits_{0}^{\infty}
        dW^{\prime} \sigma_{r.c.} (W^{\prime}) G(W^{\prime},W).
\label{SPREAD}
\end{equation}

With the currently available $\psip$ parameters
$M=3.68596$~GeV, $\Gamma_{ee}=2.19$~keV, $\Gamma_{t}=300$~keV~\cite{PDG}, 
on a collider with $\Delta=1.3$~MeV, the maximum total cross section is
640~nb; while on a collider with $\Delta=2.0$~MeV, the maximum total cross
section is 442~nb\footnote{Different accelerator has different energy
resolution, and their difference is sometimes large~\cite{jpsirep}.
Here $\Delta=1.3$~MeV corresponds to the energy resolution of
BEPC/BES~\cite{resolu} at the $\psip$ energy region; and $\Delta=2.0$~MeV
corresponds to DORIS/DASP at the same energy. Throughout this paper, these
values of the parameters are used for numerical calculation.}. 

\section{$\OP$ and $\PP$ form factors measured at $\psip$}
The decay $\psip \rightarrow \OP$ is reported by BES
to have a branching  fraction of 
$(3.8\pm1.7\pm1.1) \times 10^{-5}$~\cite{omegapi};  
and $\psip\rightarrow \pi^+\pi^-$ is reported 
by the same group to have a branching fraction of 
$(8.4\pm5.5^{+1.6}_{-3.5})\times10^{-6}$~\cite{pipi}.
With the energy resolution of BEPC/BES, these values actually  
mean that the measured cross section of $e^+e^-\rightarrow \OP$ 
at $\psi(2S)$ is $(2.4 \pm 1.3)\times10^{-2}$~nb while for 
$e^+e^-\rightarrow \pi^+\pi^-$ it is $(5.4^{+3.7}_{-4.2})\times10^{-3}$~nb.  
An earlier result of $\psip\rightarrow \pi^+\pi^-$ by DASP 
gives a branching fraction of $(8\pm5)\times10^{-5}$~\cite{DASP}.
With the energy resolution of DORIS/DASP, this means that 
the measured cross section at $\psip$ is $(3.5 \pm 2.2)\times10^{-2}$~nb.

Using these measured cross sections, together with 
Eq.~(\ref{Born}), (\ref{bexpr}), (\ref{RAD}) and (\ref{SPREAD}), 
the form factors can be estimated. In the calculation of radiative
correction, the upper limit of the integration in Eq.~(\ref{RAD}) used here
requires the knowledge of these form factors between 3.3~GeV and $\psip$
mass. For this purpose, the following $s$ dependences are assumed:
\begin{equation}
 |{\cal F}_{\OP}(s)| \propto \frac{1}{s},
\label{Fop1}
\end{equation}
or 
\begin{equation}
 |{\cal F}_{\OP}(s)| \propto \frac{1}{s^2},
\label{Fop2}
\end{equation}
for $\OP$~\cite{chernyak,Manohar,Gerard} and
\begin{equation}
 |{\cal F}_{\PP}(s)| \propto \frac{1}{s},
\label{Fpp}
\end{equation}
for $\PP$~\cite{chernyak,Manohar}.

First assume that in Eq.~(\ref{Born}), $B(s)$ is expressed 
by Eq.~(\ref{bexpr}), with the BES measured $\OP$ cross section at $\psip$, 
and normalize the $\OP$ form factor to its value at $Q^2=0$ by using the
crossed channel decay $\omega\rightarrow\gamma\pi^0$, one gets
\beqn
\lefteqn{
 {\displaystyle 
 \frac{|{\cal F}_{\OP}(m_{\psip}^2)|}{|{\cal F}_{\OP}(0)|} }} \nonumber \\
&=&\sqrt{ \displaystyle
\frac{\alpha}{3}\left(\frac{P_\gamma}{P_\omega}\right)^3
\frac{m_{\psip} \Gamma(\psip\rightarrow\omega\pi^0)}
{\Gamma(\omega\rightarrow\gamma\pi^0)
\Gamma(\psip\rightarrow\MM)}} \nonumber \\
&=& (1.6\pm0.4) \times 10^{-2} , \nonumber
\eeqn
where $P_\gamma(P_\omega)$ is the photon($\omega$) momentum in 
the $\omega(\psip)$ rest frame. 
This corresponds to the branching fraction 
$$
\BR (\psip\rightarrow\OP) = (1.6\pm0.9)\times10^{-5}~.
$$
The above results are insensitive to the $s$ dependence of the 
$\OP$ form factor: the two different $s$ dependences in 
Eq.~(\ref{Fop1}) and (\ref{Fop2}) yield the results which deviate 
from each other by only $1.1\%$ for $\OP$ form factor and $2.2\%$ 
for the branching fraction, with Eq.(\ref{Fop1}) giving 
the larger values. Such difference is well within the experimental
uncertainties of the current measurements. 

Similarly, with the BES measured $\PP$ cross section, 
$$
|{\cal F}_{\PP}(m_{\psip}^2)| = 
(4.5^{+1.5}_{-1.7}) \times 10^{-2} ,
$$
and 
$$
\BR (\psip\rightarrow\PP) = (3.5^{+2.3}_{-2.7})\times10^{-6}.
$$
With DASP result,
$$
|{\cal F}_{\PP}(m_{\psip}^2)|=0.12 \pm 0.04 ,
$$
and
$$
\BR (\psip\rightarrow\PP) = (2.6\pm1.6)\times10^{-5}.
$$
In the above equations, ${\cal B}_0$ indicates the actual branching ratio of
$\psip$ decays after continuum contribution being subtracted.

Next consider a possible phase 
between the two propagators, then instead of Eq.~(\ref{bexpr}),
Eq.~(\ref{phi}) for $B(s)$ is used in Eq.~(\ref{Born}) 
for the Born order cross section.
With an extra parameter, the form factors vary in a range, 
depending on the phase. 
Then with BES result on $\OP$, 
\beqn
\lefteqn{(1.4\pm 0.4)\times 10^{-2} \le} \nonumber \\
&&\frac{|{\cal F}_{\OP}(m_{\psip}^2)|}{|{\cal F}_{\OP}(0)|}
\le (1.8 \pm 0.5)\times 10^{-2} \nonumber
\eeqn
and 
\beqn
\lefteqn{(1.2 \pm 0.6 )\times10^{-5} \le } \nonumber \\
&& \BR (\psip\rightarrow\OP) 
\le (2.1 \pm 1.1) \times10^{-5}. \nonumber
\eeqn
The lower or upper limit herein corresponds to 
$\phi=90^\circ$ or $-90^\circ$ which
leads to maximum constructive or destructive interference between the
two propagators.

Similarly, with BES result on $\PP$, 
\beqn
\lefteqn{(3.9^{+1.3}_{-1.5})\times 10^{-2} \le} \nonumber \\
&& |{\cal F}_{\PP}(m_{\psip}^2)| 
\le (5.3^{+1.8}_{-2.1})\times 10^{-2}, \nonumber
\eeqn
and 
\beqn
\lefteqn{(2.7^{+1.8}_{-2.1}) \times10^{-6} \le} \nonumber \\
&&\BR (\psip\rightarrow\PP) \le (4.8^{+3.3}_{-3.7})\times10^{-6}.
\nonumber
\eeqn
With DASP measurement on $\PP$, 
\beqn
(0.11 \pm 0.03) \le |{\cal F}_{\PP}(m_{\psip}^2) |
\le (0.14 \pm 0.04), \nonumber
\eeqn
and 
\beqn
\lefteqn{(2.1 \pm 1.4) \times10^{-5} \le } \nonumber \\
&&\BR (\psip\rightarrow\PP) \le (3.3 \pm 2.1)\times10^{-5}.
\nonumber
\eeqn

These form factors extracted from experimentally
measured cross sections are to be compared with theoretical 
calculations. For $\OP$, a straitforward application
of Bjorken-Johnson-Low theorem for large scale time-like
momentum transfer gives~\cite{Gerard}
$$
\frac{|{\cal F}_{\OP}(s)|}{|{\cal F}_{\OP}(0)|} 
= \frac{(2\pi f_\pi)^2}{3s} = 1.66 \times 10^{-2}~~,
$$
where $f_\pi$ is the $\pi$ decay constant. This value
is in excellent agreement with the experimental measurement 
from BES with the direct one-photon annihilation contribution 
subtracted. 
On the other hand, a phenomenological model~\cite{chernyak} 
predicts
$$
  \frac{|{\cal F}_{\OP}(s)|}{|{\cal F}_{\OP}(0)|} = 
              \frac{m_{\rho}^2M_{\rho'}^2}
              {(m_{\rho}^2-s)(M_{\rho'}^2-s)},
$$
where $m_{\rho}$ and $M_{\rho'}$ are the masses of $\rho(770)$ and
$\rho(1450)$ respectively. It gives
$$
\frac{|{\cal F}_{\OP}(m^2_{\psip})|}{|{\cal F}_{\OP}(0)|}
=8.7\times 10^{-3},
$$
which deviates by two standard deviations from the above revised result 
based on BES measurement.

For $\PP$, the first order QCD calculation
relates the meson form factor with the decay constant by~\cite{brodsky} 
$$
|{\cal F}_{\PP}(s)| = 16\pi\alpha_s(s)\frac{f_\pi^2}{s}~. 
$$
Using $f_\pi=0.093$GeV and $\alpha_s(m^2_{\psip})=0.25$, one gets 
$$
|{\cal F}_{\PP}(m^2_{\psip})| = 8.0\times 10^{-3}.
$$
This is too small compared with the value extracted from the data above. 
There are also estimations by phenomenological models, e.g. 
in Ref.~\cite{fofa1}. Most of the theoretical estimations 
do not exceed~\cite{chernyak}
$$
|{\cal F}_{\PP}| \simeq \frac{0.5 \sim 0.6 \mbox{~~GeV}^2}{s}~.
$$
So the value extracted from the BES result is near the upper 
bound of the theoretical estimations. 

To explore the possible phase between the two propagators 
experimentally, it is necessary to scan the resonance shape and fit 
the $\OP$ or $\PP$ cross sections with theorectical curves. To 
accumulate sufficient
integrated luminosity at each energy point, 
this has to be done on the high luminosity colliders, such as
the upcoming CESR-c/CLEO-c~\cite{cleoc} and 
BEPC-II/BES-III~\cite{bes3}. 
  
\section{The dependence of the measurement on the 
experimental details}
In this section, the dependence of the measurement on the experimental
details is discussed. 

One important feature is that in the total measured cross section, 
the resonance part depends sensitively on the energy spread of the collider. 
The larger the energy spread, the smaller the resonance part of
the cross section. On the other hand, such energy spread 
hardly affects the one-photon annihilation part of the observed cross
section, which is a smooth function of c.m. energy $\sqrt{s}$. For example,
using Eq.~(\ref{Born}) and (\ref{bexpr}) for the Born order cross section, 
for BEPC/BES ($\Delta=1.3$ MeV) in $\EE\rightarrow\OP$ process, 
only 40.9\% of the observed cross section comes from $\psip$; 
the other 60.4\% is from the one-photon continuum and 
there is $-1.3$\% negative contribution from interference. 
For $\EE\rightarrow\PP$ process, the percentages are 41.4\%, 60.0\% and
$-1.4$\%. There are colliders with larger energy spread. In such cases the
percentage of the resonance part in the observed cross section is smaller. 
For DORIS/DASP ($\Delta=2.0$ MeV), these percentages are 32.7\%, 68.5\% and
$-1.2$\% for $\PP$.

Another important feature is that the one-photon annihilation
term, with radiative correction, depends sensitively 
on the upper limit of the integration $x_m$ in Eq.~(\ref{RAD}),
which means the invariant mass cut in the event selection. 
In contrast, the resonance part hardly changes with $x_m$ 
as long as $x_m\gg\Gamma_t/M$ due to the behavior of the Breit-Wigner formula, 
so it is virtually independent of the invariant mass cut under practical
event selection criteria. In fact, the tighter cuts on the invariant mass
of the final hadrons, which corresponds to smaller $x_m$,
the smaller the one-photon annihilation part of the observed 
cross section. 
In the calculations of this work, the value of $x_m=0.2$ is used,
which means a lower cut of $\OP$ or $\PP$ invariant mass at 
$\sqrt{1-0.2}~m_{\psip}=3.3$~GeV. Such cut or its equivalence is 
usually imposed in the event selection to 
separate the $\psip$ daughter particles from $J/\psi$'s. 

The third feature is that the treatment of the one-photon
annihilation term is sensitive to the energy on which
the data is taken. Small changes of the energy lead
to rapid variation of the resonance and the interference term.  
Experiments naturally tend to collect resonance data
at the energy which yields the maximum inclusive hadron
cross section. This energy is not the nominal resonance mass,
but somewhat higher. Nor does it necessarily
coincide with the maximum cross section of each 
exclusive mode, due to the interference effect. 
For example, with energy resolution $\Delta=1.3$~MeV,
the maximum cross sections of inclusive hadrons 
and $\OP$ mode happen at energies which are
0.14~MeV and 0.81MeV above the nominal
$\psip$ mass respectively.
At the energy which yields the maximum cross section 
of the inclusive hadrons,
the $\OP$ mode reaches only 95\% of its own maximum value; While
at the energy which yields the maximum $\OP$ cross section,
the percentages of resonance, one-photon annihilation 
and interference are 34.8\%, 57.3\% and $+7.8$\%, respectively.

\section{Conclusion}

The above analyses and estimations show that 
a large fraction of the observed cross sections of
$\EE\rightarrow\OP$ and $\EE\rightarrow\PP$ come from
the direct one-photon annihilation instead of $\psip$ decays.
This contribution must be taken into account, in order to
obtain the correct branching fractions of 
$\psip\rightarrow\OP$ and $\psip\rightarrow\PP$.
The method presented in this paper will play an important role
for the same and similar final state analysis   
in the future high luminosity experiments, like CLEO-c \cite{cleoc}
and BES-III \cite{bes3},
as the accuracy goes much higher, not only the one-photon 
annihilation part of the cross section must be treated precisely, 
the interference term could also become relevant to the 
measurements. 

\section*{Acknowledgments}
The authors are thankful for Dr. D.~H.~Zhang and Dr.~H.~B.~Li for
discussions and communications.

\end{document}